\documentclass[aps,pra,twocolumn,footinbib,
showpacs,showkeys,nobalancelastpage,
longbibliography]{revtex4-1}

\usepackage{graphicx}
\usepackage{indentfirst}
\usepackage{physics}
\usepackage{braket}
\usepackage{float}
\usepackage{amsmath}
\usepackage{amssymb}
\usepackage{verbatim}
\usepackage{epstopdf}
\usepackage{CJK}
\usepackage{esint}
\usepackage{color}
\usepackage{epsfig}
\usepackage{subfigure}
\usepackage{amsfonts}
\usepackage{footmisc}
\usepackage{scrextend}
\usepackage{multirow}
\usepackage{mathtools}
\usepackage{xr-hyper}
\usepackage[hyperfootnotes=false]{hyperref}

\usepackage[english]{babel}
\usepackage{url}
\usepackage{bm}
\definecolor{darkblue}{rgb}{0,0,0.5}
\hypersetup{
colorlinks=true,
linkcolor=black,
filecolor=black,
citecolor=darkblue,
urlcolor=black,
}
\newcommand\ceil[1]{\left\lceil{#1}\right\rceil}

\newcommand\bs[1]{\boldsymbol{#1}}

\begin{document}

\title{Mixed State Entanglement Classification using Artificial Neural Networks}
\date{\today}

\author{Cillian Harney${}^{1}$}
\author{Mauro Paternostro${}^{2}$}%
\author{Stefano Pirandola${}^{1}$}%
\affiliation{${}^{1}$Computer Science and York Centre for Quantum Technologies,University of York, York YO10 5GH, United Kingdom}
\affiliation{${}^{2}$Centre for Theoretical Atomic, Molecular and Optical Physics, School of Mathematics and Physics, Queen's University Belfast, Belfast BT7 1NN, United Kingdom}

\begin{abstract}
Reliable methods for the classification and quantification of quantum entanglement are fundamental to understanding its exploitation in quantum technologies. One such method, known as Separable Neural Network Quantum States (SNNS), employs a neural network inspired parameterisation of quantum states whose entanglement properties are explicitly programmable. Combined with generative machine learning methods, this ansatz allows for the study of very specific forms of entanglement which can be used to infer/measure entanglement properties of target quantum states. In this work, we extend the use of SNNS to mixed, multipartite states, providing a versatile and efficient tool for the investigation of intricately entangled quantum systems. We illustrate the effectiveness of our method through a number of examples, such as the computation of novel tripartite entanglement measures, and the approximation of ultimate upper bounds for qudit channel capacities.
\end{abstract}
\maketitle

The core tasks of entanglement \textit{classification} \cite{Ent_Cl,Ent_Horo_Rev, Mike_Ike} and \textit{quantification} \cite{Vedral_Meas1,Vedral_Meas2,Plenio_Meas1} are essential for future quantum technologies, and ask the seemingly straightforward questions: Given a quantum state $\rho$, is it entangled? If so, by how much is it entangled? As the system size or dimension of a quantum system grows, these questions become highly non-trivial and in general there are no universal criteria or methods to provide answers. The most popular mathematical recipe for classification, the Positive Partial Transposition (PPT) criterion (or Peres-Horodecki criterion) \cite{PPTPeres,PPTHorodecki}, applies only to $(2\otimes 2)$ or $(2\otimes 3)$ bipartite systems. As one extends to multipartite, higher dimensional quantum systems more sophisticated tools are required.\par
The application of classical machine learning tools for the study of quantum systems, such as Artificial Neural Networks (ANNs), have seen a surge of interest due to their remarkable expressive power and efficiency \cite{ML_Rev, QF_Rev, ML_Matter_Rev}. In particular, Carleo and Troyer \cite{CarleoNNS} showed that Restricted Boltzmann Machines (RBMs) offer a resoundingly appropriate classical representation of quantum states, due to their ability to perform dimensionality reduction, their non-local information distribution, and optimisation capacity \cite{RBMNNS}. Ansatzes based on this architecture are known as Neural Network Quantum States (NNS), and they have been a successful classical simulation tool in a variety of contexts such as tomography \cite{TomoNNS,MixedNNS, HomodyneNNS,RotorNNS}, open quantum system dynamics \cite{LiouvilleNNS1,LiouvilleNNS2,LiouvilleNNS3,LiouvilleNNS4,LiouvilleNNS5}, and the simulation of quantum computing \cite{QCNNS,QAOANNS,MeasNNS}.\par
The versatility of NNS also provides an excellent framework for the study of entanglement \cite{DengNNS1}. As introduced for pure, qubit states in Ref.~\cite{ECNNS}, it is possible to manipulate and constrain these neural networks in a way that guarantees a strict form of separability. These constrained variational states are known as Separable Neural Network States (SNNS). Combined with a quantum state reconstruction algorithm, this introduces a unique entanglement witness protocol based on the reconstructive performance of a SNNS with a target state.\par
 In this paper, we generalise these results to mixed, $d$-dimensional quantum states. We show how SNNS can be used to perform highly specific entanglement classification, and approximate entanglement measures to a very high degree of accuracy. The ability to implicitly characterise the space of separable states is extremely valuable, and allows one to compute entanglement measures that are otherwise extremely difficult to measure, such as the Relative Entropy of Entanglement (REE) \cite{RelEnt_Rev}.\par
This paper is structured as follows: In Section \ref{sec:NNS} we revise the NNS architecture and its variants for pure and mixed states. Section \ref{sec:SNNS} overviews separable architectures, and shows how specific forms of entanglement can be guaranteed. In Section \ref{sec:Learn_Cl_Q} the methods of classification and quantification using SNNS are discussed. Section \ref{sec:Results} provides numerical evidence for their utility through a number of relevant examples, with interesting applications in the study of noisy tripartite entanglement, bound entanglement, and quantum channel capacities. Finally, conclusions and future directions are addressed in Section \ref{sec:Conclusions}.

\section{\label{sec:NNS}Neural Network Quantum States}
\subsection{Pure states}
\begin{figure}[t!]
\centering
\hspace{-5mm}(a) NNS Qudit Architecture\\ \hspace{1cm} \\
\includegraphics[width=0.9\linewidth]{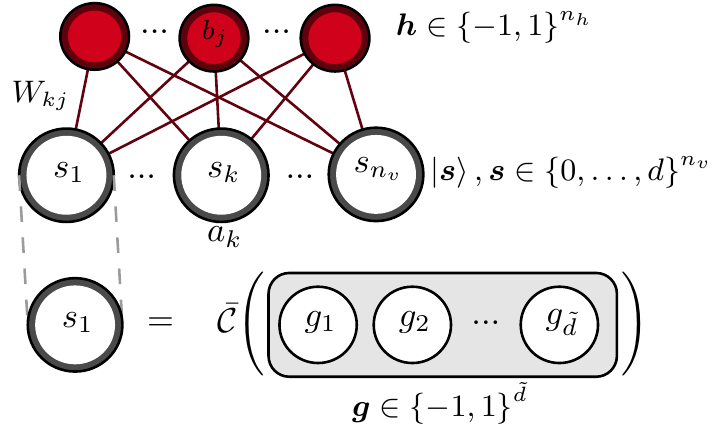}\\
\hspace{-5mm}(b) Amplitude/Phase NNS \\ \hspace{1cm} \\
\includegraphics[width=0.9\linewidth]{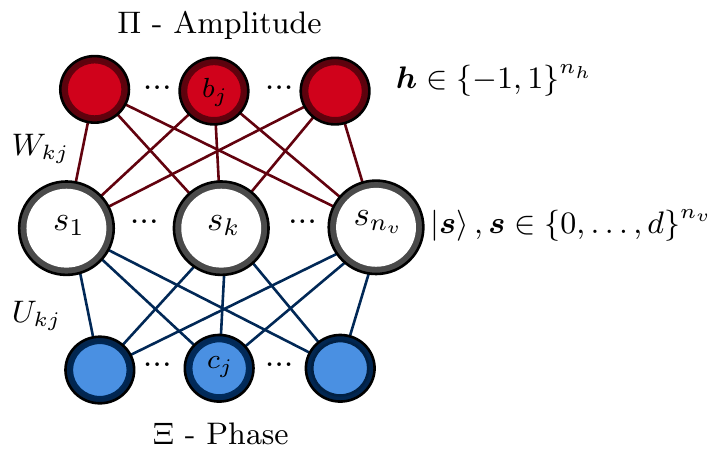}
\caption{Neural network quantum state architectures for the simulation of pure-states. Panel (a) illustrates the standard NNS construction for $n$ qudits. The visible-layer consists of $n_v\times \tilde{d}$ units which encode the accessible basis states of the target system; Here $\tilde{d}$ is the number of visible units required to encode a single qudit state where  $\mathcal{C}(\cdot)$ is some encoding function such that $\mathcal{C}(\ket{d}) = \{g_i\}_{i=1}^{\tilde{d}}$ and its inverse $\bar{\mathcal{C}}(\{g_i\}_{i=1}^{\tilde{d}}) = \ket{d}$. Correlations between qudits are captured by an $n_h$ unit hidden-layer with interconnected weights and biases. Panel (b) illustrates the amplitude/phase machine that uses two hidden-layers and only real valued parameters.}
\label{fig:QuditNNS}
\end{figure}

The simplest neural network model we can introduce is the positive, real NNS. This model uses a real valued restricted Boltzmann machine (RBM) architecture, with $n_{v}$ visible units $\bs{s} = \{s_1, \ldots, s_{n_v}\}$ representing the number of qudits being modelled within the target quantum system, fully interconnected with $n_h$ hidden units $\bs{h} = \{h_1, \ldots, h_{n_h}\}$. The visible units are typically binary valued to study $d=2$ dimensional systems, $s_i \in \{-1,1\}$ as are the hidden units $h_j \in \{-1,1\}$; however this depends on the system being modelled. This network architecture allows us to capture the correlations of the objective quantum system through network parameters:
\begin{gather}
\Pi = \{ a_k, b_j, W_{kj} \} \text{ for } k \in [1,n_v], \>\> j \in [1,n_h],\\
\bs{a} \in \mathbb{R}^{n_v}, \bs{b} \in \mathbb{R}^{n_h}, W \in \mathbb{R}^{n_v \times n_h},
\end{gather}
where $\bs{a}$ are visible biases, $\bs{b}$ are hidden biases, and $W$ is the network weight matrix. The total number of parameters is $|\Pi| = n_h\cdot n_v + n_h + n_v$ (see Fig.~\ref{fig:QuditNNS}). \par
The inherent advantage offered by the RBM architecture for generative modelling is that there are no \textit{intra}-layer connections (i.e.~there are no connections between adjacent visible units or hidden units). This allows for an ansatz that is independent from the activations of the hidden state space. Thus, one can define a positive NNS wavefunction as \cite{CarleoNNS}
\begin{align}
\Psi_\Pi(\bs{s}) = e^{\sum_{k=1}^{n_v} a_k s_k} \prod_{j=1}^{n_h} 2\cosh\left( \sum_{k} W_{kj} s_k + b_j \right),
\end{align}
and therefore the NNS is $ \ket{\Psi_\Pi} = \sum_{\bs{s}} \Psi_{\Pi}(\bs{s}) \ket{\bs{s}}$.\par
Whilst NNS have typically been applied to qubit systems using binary visible units, one can extend the modelling to $d$-dimensional qudits by using a set of visible binary neurons that collectively represent a single qudit \cite{RotorNNS}. One may choose to encode $d$-dimensional states using a collection of $\tilde{d}$ visible, binary neurons via an encoding function $\mathcal{C}$, i.e.~
\begin{equation} 
\ket{s} \mapsto \mathcal{C}(s) =  \{g_1, g_2, \ldots, g_{\tilde{d}}\} = \bs{g}.
\end{equation}
The $n_v$ qudit visible-layer can then be encoded into $\tilde{n}_v = \tilde{d} n_v  > n_v$ visible neurons,
\begin{equation}
\bs{s} = \{s_1,s_2,\ldots,s_{n_v}\} \mapsto \{ \bs{g}_1, \bs{g}_2, \ldots , \bs{g}_{\tilde{n}_v}\}.
\end{equation}
We may identically define the qudit decoding function $\bar{\mathcal{C}}$ such that $\bar{\mathcal{C}}(\bs{g}) = \ket{s}$. One may encode qudits into binary codes on the visible-layer $\ket{s} \mapsto \text{bin}(s)$, requiring $\tilde{n}_v = \ceil{\log_2 d} n_v$ visible binary neurons, which however requires $d = 2^r$ for some integer $r$ in order to admit a complete basis set. For arbitrary $d$ it may be more useful to utilise one-hot encoding such that  $\ket{s} \mapsto \text{onehot}(s) = \bs{e}_s^{d}$ where $\bs{e}_s^{d}$ is a $d$-length vector that is zero at all indices except index $s$.\par

In order to study non-positive quantum states one can introduce complex network parameters. Letting $a_k = \alpha_k + i\beta_k$, $b_j = \gamma_j + i\lambda_j$, and $W_{kj} = \Gamma_{kj} + i\Lambda_{kj}$, then the NNS wavefunction is
\begin{gather}
\Psi_{\Pi} (\bs{s})  =  e^{ \sum_{k=1}^{n_v} (\alpha_k + i\beta_k) s_k}  \prod_{j=1}^{n_h} 2\cosh\left( \theta_j^\gamma + i \theta_j^\lambda \right) ,
\end{gather}
where $\theta_j^{\gamma} = \sum_k \Gamma_{kj} s_k + \gamma_j,$ and $\theta_j^\lambda = \sum_k \Lambda_{kj} s_k + \lambda_j$.
Thus the NNS can exhibit phase properties of quantum states. The network parameter set extends to $\Pi = \{ a_k, b_j, W_{kj}\} \in \mathbb{C}$.\par
Alternatively one can preserve reality of network parameters by restructuring the nature of the NNS ansatz itself. In particular we can construct an ansatz that uses two RBMs that unify to represent a complete state. Defining a variational phase state $\Phi_{\Xi}(\bs{s})$, and amplitude state $\Psi_{\Pi}(\bs{s})$, this network ansatz is given as \cite{TomoNNS}
\begin{align}
\ket{\Psi_{\Pi,\Xi}} &= \sum_{\bs{s}} e^{i \log{\Phi_{\Xi}(\bs{s})}} \Psi_{\Pi}(\bs{s}) \ket{\bs{s}} \label{eq:CPansatz}.
\end{align}
Therefore both the variational phase and amplitude networks need only be real valued, since the complex/phase properties of the state are managed through the complex exponential. The state is now defined by two parameter sets, $\Pi = \{ a_k, b_j, W_{kj}\} \in \mathbb{R}$ and $\Xi = \{ c_k, d_j, U_{kj}\} \in \mathbb{R}$.

\subsection{Mixed States\label{sec:MixedStates}}
To extend the variational ansatz to mixed states requires the addition of a hidden mixing-layer with $n_m$ hidden units, capable of encoding the classical probability distribution of the mixed quantum state \cite{LiouvilleNNS2,LiouvilleNNS3,LiouvilleNNS4}. The network state can be constructed from two sets of variational network parameters: $\Pi = \{ c_p, U_{kp} \}$, $c_p \in \mathbb{R}^{n_m}$ and $U_{kp}\in\mathbb{C}^{n_v\times n_m}$ encoding the mixing probabilities \footnote{The mixing hidden biases $c_p$ are set as real, while the network weights $U_{kp}$ are complex. This is just a simplification, since imaginary components of $c_p$ are negated in the network output functions. If $c_p$ were complex, it would be clear in Eq.~(\ref{eq:phi_lab}) that only the real components would be utilised, since $\phi_p({\bs{\alpha}, \bs{\beta}}) =   c_p + c_p^*+ \sum_k U_{kp} \alpha_k + U_{kp}^* \beta_k = 2\text{Re}(c_p) + \sum_k U_{kp} \alpha_k + U_{kp}^* \beta_k $.} and the previously defined $\Xi = \{a_k, b_j, W_{kj}\} \in \mathbb{C}$ which encodes the pure-state probability distribution. Let the density-matrix row and column degrees of freedom be described by basis vectors $\{\bs{\alpha},\bs{\beta}\}$ respectively. As these parameter sets are independent, we may describe a density-matrix element as a contribution from a classical mixing state $\mathcal{P}_{\Pi}$ and a pure-state $\sigma_{\Xi}$. The contribution from a classical mixing network is given by
\begin{gather}
\mathcal{P}_{\Pi}^{\bs{\alpha}, \bs{\beta}}  = \prod_{p=1}^{n_m} \cosh{( \phi_p({\bs{\alpha},\bs{\beta}} ) )}, \\
\phi_p({\bs{\alpha}, \bs{\beta}}) =   c_p + \sum_k U_{kp} \alpha_k + U_{kp}^* \beta_k. \label{eq:phi_lab}
\end{gather} 
where $x^*$ denotes complex conjugation. Meanwhile the pure-state contribution is
\begin{gather}
\sigma_{\Xi}^{\bs{\alpha}, \bs{\beta}} = e^{{\omega}({\bs{\alpha}, \bs{\beta}})} \prod_{j=1}^{n_h}\cosh{( \theta_j (\bs{\alpha}))}  \cosh{( \theta_{j}^*(\bs{\beta}) )}, \\
{\omega}({\bs{\alpha}, \bs{\beta}}) = \sum_k a_k \alpha_k + a_k^* \beta_k, \label{eq:omega_lab}\\
\theta_j(\bs{x}) = b_j + \sum_{k} W_{kj} x_k . \label{eq:theta_lab}
\end{gather}\par
\begin{figure}[t!]
\includegraphics[width=\linewidth]{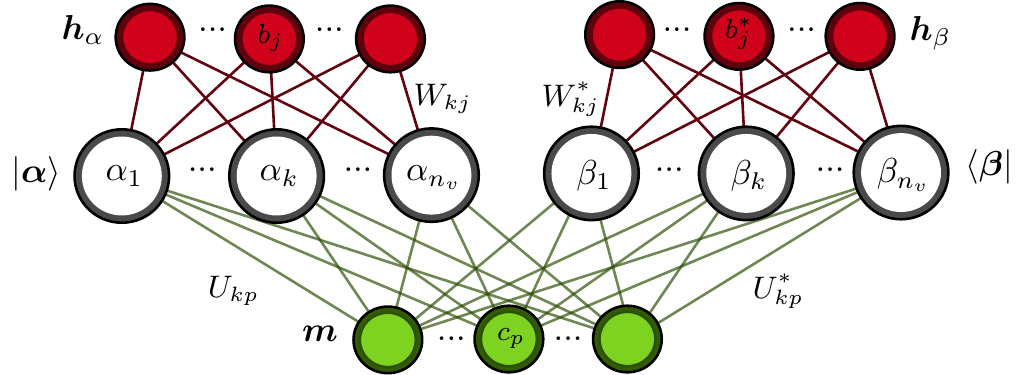}\\
\caption{A restricted Boltzmann machine architecture for the simulation of (generally entangled) density matrices using complex parameters. }
\label{fig:MixedNNS}
\end{figure}

The complete variational state can therefore be constructed as a sum over all density-matrix elements, 
\begin{equation}
\rho_{\Pi,\Xi} =  \sum_{\bs{\alpha}, \bs{\beta}} \sigma_{\Xi}^{\bs{\alpha}, \bs{\beta}} \cdot \mathcal{P}_{\Pi}^{\bs{\alpha}, \bs{\beta}} \ket{\bs{\alpha}}\!\bra{\bs{\beta}}= {\mathcal{P}}_\Pi \odot {\sigma_\Xi} \label{eq:NDM},
\end{equation}
where $\odot$ is the Hadamard product. It is important to emphasise that the classical mixing state ${\mathcal{P}}_{\Pi}$ cannot capture quantum correlations, only classical correlations. Hence the pure-state ${\sigma}_{\Xi}$ alone simulates the quantum correlations within the network state. This architecture is presented in Fig.~\ref{fig:MixedNNS}.\par
The network parameters in this ansatz are necessarily complex, but one can create a reformulated ansatz in order to use only real parameters. One \textit{could} use the NNS used in Eq.~(\ref{eq:CPansatz}) to learn a vectorised density-matrix $\bs{\rho} = \ket{\rho_{\Pi,\Xi}}$. Whilst optimal convergence towards the target vectorised mixed state is possible in this way, the ansatz itself is neither Hermitian or positive semi-definite under reshaping to a density-matrix,  i.e.~$\rho = \text{vec}^{-1}(\bs{\rho})$  is not a valid density-matrix. \par
Instead one can restructure the mixed state ansatz in order to take a closer form to the complex exponential format utilised in the previous sections. Let the real parameter sets $\Xi, \Pi$ be used to describe the pure-state phase and amplitude networks respectively, and the complex parameter set $\Omega$ used to describe the mixing network. Recall a pure state wavefunction in complex exponential form $\Psi_{\Pi,\Xi}(\bs{\alpha}) = e^{i\log \varphi_{\Xi} (\bs{\alpha})} \sigma_{\Pi} (\bs{\alpha})$. It is useful to define the following functions of our pure density-matrix phase/amplitude wavefunctions
\begin{align}
&\Phi_{\Xi}^{\bs{\alpha},\bs{\beta}} = \frac{\varphi_{\Xi} (\bs{\alpha})}{\varphi_\Xi (\bs{\beta})}, \>\>\>\> \Gamma_{\Pi}^{\bs{\alpha},\bs{\beta}} = \sigma_{\Pi} (\bs{\alpha})\sigma_\Pi (\bs{\beta}). \label{eq:PureDecomp}
\end{align}
In order to incorporate the classical mixing we need a mixing-layer that takes a similar vectorised form. Omitting the visible biases which are already possessed by the pure-states, the mixing-layer takes the form
\begin{gather}
\mathcal{P}_{\Omega}^{\bs{\alpha},\bs{\beta}} = \prod_{p=1}^{n_m} \cosh ( \mu_p + i\psi_p)  = \prod_{p=1}^{n_m} r_p^{\bs{\alpha},\bs{\beta}} e^{i\log \vartheta_p^{\bs{\alpha},\bs{\beta}}}, \\
\mu_p({\bs{\alpha}, \bs{\beta}})  = c_p + \sum_k R_{kp} (\alpha_k + \beta_k), \\
\psi_p({\bs{\alpha}, \bs{\beta}})  =   \sum_k I_{kp} (\alpha_k - \beta_k),
\end{gather}
where $R_{kp} = \text{Re}(U_{kp})$ and $I_{kp} = \text{Im}(U_{kp})$ denote the real and imaginary components of the mixing network respectively. One can then construct the following phase and amplitude functions for the classical mixing 
\begin{align}
r_{\Omega}^{\bs{\alpha},\bs{\beta}} &= \prod_{p=1}^{n_m} \sqrt{\cosh(\mu_p +  i\psi_p)\cosh(\mu_p - i\psi_p)}, \\
\vartheta_{\Omega}^{\bs{\alpha},\bs{\beta}} &= \prod_{p=1}^{n_m}\exp\left[{\frac{1}{2i} \log \left( \frac{-\cosh(\mu_p + i\psi_p)}{\cosh(\mu_p - i\psi_p)}\right)}\right],
\end{align}
such that the vectorised mixing state takes the form $e^{i \log \ket{\vartheta_{\Omega}}} \ket{r_{\Omega}}$. This allows for any element of the complete mixed state to be expressed according to
\begin{align}
\rho_{\text{$\Omega,\Pi,\Xi$}}^{\bs{\alpha},\bs{\beta}} = e^ {i\log \left({\Phi_{\Xi}^{\bs{\alpha},\bs{\beta}} \vartheta_{\Omega}^{\bs{\alpha},\bs{\beta}}}\right)} \Gamma_{\Pi}^{\bs{\alpha},\bs{\beta}} r_\Omega^{\bs{\alpha},\bs{\beta}}. \label{eq:vecNDM}
\end{align}

\section{\label{sec:SNNS}Separable Neural Network Architectures}

\subsection{Separable Pure Network States}
Through restrictions on the connectivity of the weight matrix $W_{kj}$, one can guarantee separability of the generative network state. Let us define $\mathcal{K}$ as a collection of $K$-disjoint subsets $\mathcal{K} = \{\bs{k}_l\}_{l=1}^{K}$, that collect qudit indices from an $n$-qudit system. More precisely,
\begin{gather}
\mathcal{K} = \bigcup_{l=1}^{K} \bs{k}_l, \text{ s.t } \{1,\ldots,n\} \subseteq \mathcal{K}, \label{eq:all_modes}\\
\bs{k}_m \cap \bs{k}_l = \varnothing, \> \forall ~m\neq l \in \{1,\ldots,n\}. \label{eq:modes_disjoint}
\end{gather}
In Eq.~(\ref{eq:all_modes}) we have demanded that the global partition set necessarily contains all $n$-qudits in the system, and that subsets of qudits are disjoint in Eq.~(\ref{eq:modes_disjoint}). Hence, an $n$-qudit, pure-state $\ket{\Psi}$ is defined to be $\mathcal{K}$-separable if it can expressed as a tensor-product of sub-states $\ket{\Psi} = \bigotimes_{\bs{k} \in \mathcal{K}} \ket{ \psi_{\bs{k}}}$, i.e.~it is separable with respect to the partition set $\mathcal{K}$. This is a very precise format of separability, as it precisely specifies the arrangement of entangled parties. If we were to disregard specific party orderings we would refer to $(|\mathcal{K}| = K)$-separability.\par
Disjointedness in this definition of $\mathcal{K}$-separability ensures that each qudit is only entangled with respect to a single subset of the quantum system. This provides a specific level of detail to the entanglement structure, while also degenerating many forms of entanglement that we may not be interested in. For example, genuine tripartite entanglement under disjoint $\mathcal{K}$-separability allows for only a single set $\mathcal{K} = \{\bs{k}_1\} = \{1,2,3\}$ with no partitions. We may then define \textit{non-disjoint} $\mathcal{K}$-separability as an extension of the previous definition simply by removing the conditions in Eq.~(\ref{eq:modes_disjoint}). Using this non-disjoint definition, genuine tripartite entanglement allows for many more definitions,
$\mathcal{K} = \{1,2,3\}, \{1,2|2,3\}, \{1,2|2,3|1,3\}, \ldots $, which is studied in later sections.\par
To strictly impose either type of separability on an NNS, the goal is to express the wavefunction of the network state in the following form
\begin{equation}
\Psi_{\Pi}(\bs{s}) = \prod_{l=1}^K \psi_{\Pi}^{\bs{k}_l}\left(\bs{s}\right),
\end{equation}
where $\psi_{\Pi}^{\bs{k}_l}$ are separable sub-wavefunctions that describe the behaviour of qudits in the partition $\bs{k}_l$. We may then construct an analogous hidden-layer partition set $\mathcal{H} = \{\bs{h}_l\}_{l=1}^{K}$, which assigns a subset of hidden units to each visible subset of entangled qudits $\mathcal{K} = \{\bs{k}_l\}_{l=1}^{K}$.
By segmenting the layer of hidden units into these $K$-subsets and applying the following restriction to the weight matrix
\begin{equation}
W_{ij} = 0 \text{ for } i \in \bs{k}_l, \>\> j \notin \bs{h}_l, \>\> \forall~l\in \{1,\ldots,K\} \label{SepCond},
\end{equation}
this condition then provides the complete, $\mathcal{K}$-separable network state
\begin{gather}
{\Psi_{\Pi|\mathcal{K}} ({\bs{s}})} = \prod_{l=1}^{K}  e^{\tilde{\omega}_l ({\bs{s}})} \prod_{j\in\bs{h}_{l}} 2\cosh\left( \theta_l^j ({\bs{s}}) \right) \nonumber ,\\
\theta_l^j(\bs{s}) = \sum_{i\in\bs{k}_l} W_{ij} s_i + b_j, \>\> \tilde{\omega}_l(\bs{s}) = \sum_{i\in \bs{k}_l} a_i s_i.
\end{gather}

\begin{figure}[t!]
(a) GHZ-type entanglement \\ \hspace{1cm} \\
\includegraphics[width=0.7\linewidth]{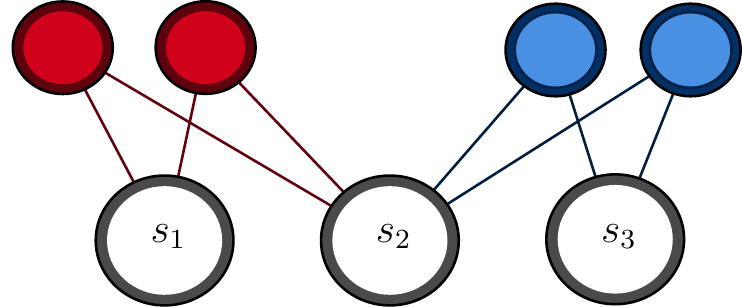}\\ \hspace{1cm} \\
(b) W-type entanglement\\ \hspace{1cm} \\ 
\includegraphics[width=0.7\linewidth]{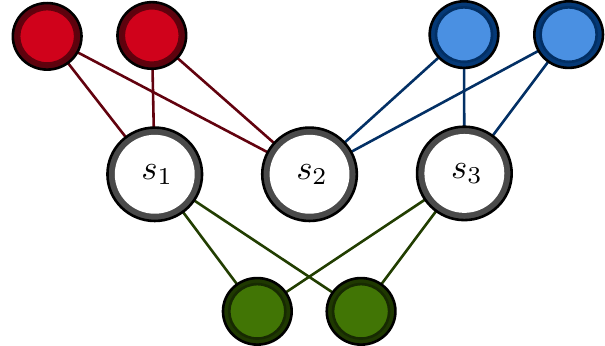}
\caption{Different pure-state network architectures used to simulate genuine tripartite entanglement. Panel (a) depicts a form of GHZ-type entanglement according to the partition set $\mathcal{K}_{\text{GHZ}} = \{1,2|2,3\}$. Notice that qudits 1 and 3 do not possess a direct connection, but may relay correlations through qudit 2. Panel (b) illustrates a non-disjoint, W-type entanglement structure according to $\mathcal{K} = \{1,2|2,3|1,3\}$. }
\label{fig:SSepndm}
\end{figure}

\subsection{Separable Neural Network Density Matrices}
Whilst pure-states are $\mathcal{K}$-separable when they can be expressed as the tensor product of $|\mathcal{K}| = k$ local sub-states, a mixed state possesses a form of separability iff it can be expressed as a convex combination of local sub-states ${\rho}^{\{\bs{k}_l\}_{l=1}^{K}}$. It is now useful to define two distinct forms of separability; \textit{consistent} and \textit{inconsistent} mixed-multipartite separability. \par
A state is consistently $\mathcal{K}$-separable if it can be expressed as a convex combination of states which \textit{all} admit an identical form of separability, 
\begin{equation}
{\rho}^{\mathcal{K}} = \sum_j p_j \bigotimes_{\bs{k}\in\mathcal{K}} {\rho}_j^{\bs{k}}.
\end{equation}
On the contrary, a state is inconsistently $\{\mathcal{K}_j\}$-separable if it is a mixture of states with different entanglement properties, 
\begin{equation}
{\rho}^{\{\mathcal{K}_j\}} = \sum_j p_j \bigotimes_{\bs{k}\in\mathcal{K}_j} {\rho}_j^{\bs{k}},
\end{equation}
so its entanglement properties are defined by a combination of constituent $\mathcal{K}_j$-separabilities. Precise classification methods are much more difficult for mixed states, however there are still some very useful approaches that can be introduced using NNS.\par
Consistently $\mathcal{K}$-separable states require a direct application of the separability conditions given by Eq.~(\ref{SepCond}) onto the pure-state of the NNS. Since the mixing state cannot capture quantum correlations, it is already separable and requires no restrictions. It is thus expedient to apply the separability conditions of Eq.~(\ref{SepCond}) onto the pure-states of the mixed NNS, restricting the capacity of the neural network to simulate quantum correlations. Enforcing separability on the pure density-matrix in this way
\begin{gather}
\sigma_{\Xi|\mathcal{K}}^{{\bs{\alpha},\bs{\beta}}} =  \prod_{l=1}^{K} e^{{\omega}_l(\bs{\alpha},\bs{\beta})}  \prod_{j\in\bs{h}_l} \cosh(  \theta_l^j(\bs{\alpha}) ) 
\cosh( {{\theta_l^{j*}}(\bs{\beta})} ) , \nonumber \\
{\omega}_l(\bs{\alpha},\bs{\beta}) = \sum_{i\in \bs{k}_l} a_i \alpha_i + a_i^{*} \beta_i,
\end{gather}
thus provides a NNS guaranteed to be consistently $\mathcal{K}$-separable
\begin{equation}
\rho_{\Pi,\Xi}^{\mathcal{K}}= {\mathcal{P}}_{\Pi} \odot \sigma_{\Xi|\mathcal{K}}.
\end{equation}\par
If one wishes to enforce complete separability such that for an $n$-qudit state ${\rho} = \sum_j p_j \bigotimes_{m=1}^{n} {\rho}_j^m$, one can of course just apply consistent separability onto the network state via the separability set $\mathcal{K} = \{1|2|,\ldots,|n\}$ in an identical manner as before. However, as the state is completely separable, there are no quantum correlations and the pure-states in the network ansatz are not necessary for simulation of the state. It can then be simplified to ${\rho}_{\Pi} = {\mathcal{P}}_{\Pi}$, and we can simulate completely separable mixed quantum systems using an RBM with a classical mixing-layer only \footnote{We are also free to combine visible layer biases $\bs{a}$ into the parameter set $\Pi$, since they are inherently local and thus invoke separable correlations.}
\begin{align}
{\rho}_{\Pi}^{\text{Sep}} &= \sum_{\bs{\alpha},\bs{\beta}} e^{ {\omega}(\bs{\alpha},\bs{\beta})}  \prod_{p=1}^{n_m} \cosh{( \phi_p(\bs{\alpha},\bs{\beta}) )} \ket{\bs{\alpha}}\!\bra{\bs{\beta}} \label{eq:FSep}.
\end{align}\par
Unfortunately, it is not possible to \textit{strictly} classify an inconsistently separable mixed state according to ansatzes discussed in this Section. Take the tripartite example
\begin{equation}
{\rho} = \sum_{j} p_j {\rho}_j^{\{1,2|3\}} +  \sum_{k} p_k {\rho}_k^{\{1|2,3\}}  + \sum_{m} p_m {\rho}_m^{\{1,3|2\}},\label{eq:Cheap}
\end{equation}
which can be thought of as ``cheap" genuine tripartite entangled state. We can certainly define an NNS that can reconstruct a state of this form (trivially, one can utilise a fully connected NNS that can reconstruct $\rho$); however we cannot specify all three forms of separability in $\rho_{}$ without also allowing the NNS to potentially manifest genuine, pure tripartite entanglement. One can instead utilise independent consistently separable NNS according to the partitions $\{1,2|3\}$, $\{1,3|2\}$ and $\{2,3|1\}$ in order to quantify the amount of entanglement in the target state with respect to each partition.

\section{\label{sec:Learn_Cl_Q}Classifying and Quantifying Entanglement}
\subsection{Learning of Quantum States}
We present a learning protocol for a pure NNS $\ket{\Psi_{\Pi,\Xi}}$ to reconstruct a target state $\ket{\varphi}$ using the ansatz from Eq.~(\ref{eq:CPansatz}), which is then extendible to mixed states. We employ a unified learning approach, where the variational state optimises the global, vectorised fidelity with a target state, rather than separate phase and amplitude fidelities. We may define the loss function as the negative logarithmic fidelity between two pure-states as a function of our set of variational parameters
\begin{align}
\mathcal{L}  = -  \log \sqrt{\frac{|\braket{\Psi_{\Pi,\Xi} | \varphi}|^2}{\braket{\Psi_{\Pi,\Xi} |\Psi_{\Pi,\Xi}} \braket{\varphi | \varphi}}}.
\end{align}
Splitting these wavefunctions into respective phase and amplitude functions, 
\begin{equation}
{\Psi_{\Pi,\Xi}}({\bs{s}})= {\psi_{\Pi}}({\bs{s}}) \>e^{ i \log({\phi_{\Xi}}({\bs{s}}) ) }, \> \varphi({\bs{s}}) = \lambda({\bs{s}})\> e^{i \log(\xi({\bs{s}}) )},
\end{equation}
we wish to compute the derivatives of the unified cost function with respect to the parameter sets $\{\Pi,\Xi\}$. Since these wavefunctions utilise only real parameters, it is expedient to compute the derivatives using the following chain rule formulation,
\begin{equation}
\nabla_{k}^{\psi_\Pi} \mathcal{L} = \frac{\partial \mathcal{L}}{\partial \ket{\psi_\Pi}} \cdot \frac{\partial \ket{\psi_\Pi}}{\partial \Pi_k}, \>\>\>\> \nabla_{k}^{\phi_\Xi} \mathcal{L} = \frac{\partial \mathcal{L}}{\partial \ket{\phi_\Xi}} \cdot \frac{\partial \ket{\phi_\Xi}}{\partial \Xi_k}.
\end{equation}
Computing these gradients will provide the necessary parameter update rules at the $m^{\text{th}}$ iteration to the $k^{\text{th}}$ network parameter by gradient descent, taking the form
\begin{align}
\Pi_{k}^{m+1} = \Pi_{k}^{m} - \eta \nabla_{k}^{\psi_\Pi} \mathcal{L},\>\>\>\>\>\>
\Xi_{k}^{m+1} = \Xi_{k}^{m} - \eta \nabla_{k}^{\phi_\Xi} \mathcal{L} \label{UpdatePhase},
\end{align}
where $\eta$ is some learning rate small enough such that the network state converges to the target state over sufficient iterations of the learning scheme.\par
Defining the quantity
\begin{equation}
\Delta(\bs{s}) =  {{\braket{\Psi_{\Pi,\Xi}|\varphi}}}^{-1} {e^{i \log \frac{{\phi_{\Xi}}({\bs{s}})}{\xi({\bs{s}})}}},
\end{equation}
complete gradients with respect to variational parameters can therefore be computed as
\begin{align}
&\nabla_k^{\psi_\Pi}\mathcal{L} = \sum_{\bs{s}} \Bigg[ \frac{{\psi_\Pi}({\bs{s}})}{|\Psi_{\Pi,\Xi}|^2} - \lambda({\bs{s}}) \text{Re}\big[\Delta(\bs{s})
\big] \Bigg]  {\mathcal{O}}_{k}^{\Pi} \ket{\psi_\Pi}, \\
&\nabla_k^{\phi_\Xi}\mathcal{L} = -\sum_{\bs{s}} \Bigg[ \frac{\lambda({\bs{s}})  {\psi_\Pi}({\bs{s}})}{{\phi_\Xi}({\bs{s}})} \text{Im}\big[
\Delta(\bs{s})
\big]\Bigg] {\mathcal{O}}_{k}^{\Xi} \ket{\phi_\Xi}, \label{eq:Gradients}
\end{align}
where ${\mathcal{O}}_{k}^{\Pi} = \text{diag}\left({\partial_{\Pi_k} \log \ket{\psi_\Pi}}\right)$, ${\mathcal{O}}_{k}^{\Xi} = \text{diag}\left({\partial_{\Xi_k} \log \ket{\phi_\Xi}}\right)$ denote diagonal matrices containing the logarithmic derivatives of the network state with respect to the $k^\text{th}$ amplitude and phase network parameters respectively. Utilising Eq.~(\ref{eq:Gradients}) in the update rule given by Eq.~(\ref{UpdatePhase}), the phase and amplitude properties will optimise in a unified manner, maximising the fidelity between the network and the target state endowed with non-trivial phase structure.\par
Fortunately this learning procedure is readily extended to mixed states via the ansatz in Eq.~(\ref{eq:vecNDM}). Since the variational state is in a complex exponential format, one then formulates a cost function based on the fidelity between the vectorised density-matrix and the vectorised target state. The extension is straightforward and explained in Appendix \ref{sec:DerivsULM}. \par
As shown in Ref.~\cite{ECNNS} separable neural network states can be used to perform entanglement classification and provide entanglement measures of pure, two-dimensional quantum states. Using qudit sub-encoding and the mixed state architectures discussed in the previous sections, these ideas can be extended to classification of more complex quantum systems.\par
Let us devise a precise decision rule for classification. Consider a target $n$-qudit state $\sigma$, a $\mathcal{K}$-separable learner $\rho_{\Omega}^{\mathcal{K}}$, and a free, entangled learner $\rho_{\Omega}^{\text{Ent}}$ which have both been optimised with respect to reconstructing $\sigma$. Using the Bures fidelity, $F(\sigma, \rho) =  \text{Tr}\sqrt{\sqrt{\sigma} \rho \sqrt{\sigma}}$, we denote the \textit{reconstruction fidelity} of a learning process as the final/optimal fidelity achieved after a given number of learning iterations. A target $\sigma$ is learnable via $\rho_{\Omega}^{\text{Ent}}$ iff its reconstruction fidelity satisfies
\begin{equation}
F(\sigma, \rho_{\Omega}^{\text{Ent}}) \geq F_{\text{opt}} = 1 - \epsilon,
\end{equation}
for a sufficiently small threshold $\epsilon$. The choice of $F_{\text{opt}}$ determines the reliability of classification, and in our numerical experiments we fix $\epsilon \leq 10^{-4}$. The accuracy of this reconstruction via free learning also benchmarks the satisfactory computational resources required in the network, informing the separable reconstruction. \par
One can reliably infer that a target state is $\mathcal{K}$-separable if it is learnable by both a free NNS ($\rho_{\Omega}^{\text{Ent}}$), and a $\mathcal{K}$-separable NNS ($\rho_{\Omega}^{\mathcal{K}}$). Then the NNS reconstruction fidelities must satisfy
\begin{equation}
F(\sigma, \rho_{\Omega}^{\mathcal{K}}) \geq  F(\sigma, \rho_{\Omega}^{\text{Ent}}) \geq F_{\text{opt}}.
\end{equation}
Otherwise, the state is entangled to a higher degree. One may then quantify the entanglement content of the target by investigating the distance between $\sigma$ and an approximation to the closest $\mathcal{K}$-separable state.

\subsection{Quantifying Entanglement}
The most difficult aspect of quantifying entanglement stems from the complicated nature of characterising the space of separable quantum states. Thanks to the implicit guarantee of specific separability, SNNS offer an extremely useful tool to help with this, and provide the opportunity to study a variety of entanglement measures that are otherwise much too difficult to explore. \par
Let us consider measures $E$ that satisfy the general properties of a valid entanglement measure \cite{EntMeasure}. Many important types of $E$ are constructed as a geometric optimisation problem with respect to the space of all fully separable states  $\mathcal{D}_{\text{Sep}}$. That is, given a target state $\sigma$ and a distance measure (possibly quasi-distance measure) $f$,
\begin{gather}
E(\sigma) = \min_{\rho \in \mathcal{D}_{\text{Sep}}}\hspace{-2mm}f(\sigma, \rho), \label{eq:GeoEM}\\
\text{if } \sigma \in\mathcal{D}_{\text{Sep}} \implies E(\sigma) = 0,\\
\text{if } \sigma \notin \mathcal{D}_{\text{Sep}} \implies E(\sigma) > 0.
\end{gather}
These are entanglement measures which are computed by locating the Closest Separable State (CSS) $\sigma^{\star}$ to $\sigma$, with respect to the distance measure $f$. For such measures, the employment of SNNS to parameterise the separable states $\rho_{\Omega} \in \mathcal{D}_{\text{Sep}}$ is extremely useful, as it offers an efficient way to perform this optimisation. Furthermore, since SNNS are inherently separable, they will always approximate an upper bound on $E$, since they are certifiably limited in the quantum correlations that they are able to simulate. This is,
\begin{equation}
E(\sigma) \leq E_{\Omega}(\sigma) = \min_{\rho_{\Omega} \in \mathcal{D}_{\text{Sep}}}\hspace{-2mm}f(\sigma, \rho_{\Omega}).
\end{equation}
\par
To generalise, we may construct a measure $E^{\mathcal{K}}$ which is analogous to $E$, but is defined with respect to the space of all states which are at most $\mathcal{K}$-separable. Defining the set of all states that are $\mathcal{K}$-separable as 
$\mathcal{D}_{\mathcal{K}}$, then the set of all states that are at most $\mathcal{K}$-separable is given by \footnote{Note that in Eq.~(\ref{eq:SSepMost}) the union runs over all $|\mathcal{K}^{\prime}| > |\mathcal{K}|$. This is necessarily a strict inequality. Suppose $|\mathcal{K}| = k$ such that $\mathcal{K}$ describes a form of $k$-separability. The set of all $\mathcal{K}$-separable states thus inherits all states which are $(l < k)$-separable, but there are other forms of $|\mathcal{K}^{\prime}| =k$ separability which it will not inherit. This is true for disjoint and non-disjoint constructions, however in some non-disjoint cases these sets coincide.}
\begin{equation}
\tilde{\mathcal{D}}_{\mathcal{K}} = \mathcal{D}_{\mathcal{K}} \bigcup_{|\mathcal{K}^{\prime}| > |\mathcal{K}|} \mathcal{D}_{\mathcal{K}^{\prime}}. \label{eq:SSepMost}
\end{equation}
Assuming a measure of the form Eq.~(\ref{eq:GeoEM}), then we can define
\begin{gather}
E^{\mathcal{K}}(\sigma) = \min_{\rho \in \tilde{\mathcal{D}}_{\mathcal{K}}}\hspace{-1mm}f(\sigma, \rho) \leq E_{\Omega}^{\mathcal{K}}(\sigma), \\
\text{if } \sigma \in \tilde{\mathcal{D}}_{\mathcal{K}} \implies E^{\mathcal{K}}(\sigma) = 0,\\
\text{if } \sigma \notin \tilde{\mathcal{D}}_{\mathcal{K}} \implies E^{\mathcal{K}}(\sigma) > 0.
\end{gather} 
$E^{\mathcal{K}}$ satisfies all the general properties of an entanglement measure, but now with respect to $\tilde{\mathcal{D}}_{\mathcal{K}}$, and is therefore able to classify/quantify more complex forms of entanglement.\par
Let us specify some important entanglement measures which SNNS can utilise, starting from the Geometric Measure of Entanglement (GME) \cite{GME}. For pure-states, the GME is the maximum fidelity that can be obtained between a target state $\ket{\sigma}$ and the set of pure, at most $\mathcal{K}$-separable states $\tilde{\mathcal{B}}_{\mathcal{K}}$
\begin{equation}
E_{\text{G}}(\ket{\sigma}) = \max_{\ket{\varphi} \in \tilde{\mathcal{B}}_{\mathcal{K}}} \hspace{-1mm} F(\ket{\sigma},\ket{\varphi}).  \label{eq:GME}
\end{equation}
For more sophisticated mixed state approaches, it is expedient to employ any number of density-matrix distance measures. Several important examples include the trace distance
\begin{equation}
E_{C_1}(\sigma) =  \frac{1}{2} \min_{\rho \in \mathcal{D}_{\text{Sep}}}\hspace{-1.5mm} \| \sigma- \rho\|_1,
\end{equation}
where $\| X\|_1 = {\text{Tr} \sqrt{X^{\dagger} X}}$ or the Bures metric
\begin{equation}
E_{B}(\sigma) = \min_{\sigma \in \mathcal{D}_{\text{Sep}}} \hspace{-2mm}\left[ 1 - F^2(\rho,\sigma) \right],
\end{equation}
where $F$ is the Bures fidelity as before. These quantities are readily approximated via SNNS, and easily specified to different forms of $\mathcal{K}$-separability.\par
Of particular interest is the Relative Entropy of Entanglement (REE) \cite{RelEnt_Rev}, an entanglement measure that has many applications in quantum communications and channel capacities \cite{PLOB}. The REE is based on the quantum relative entropy (QRE), a kind of distance measure between two quantum states where
\begin{equation}
S({\rho}\|{\sigma}) = \text{Tr}\left[ {\rho}\left(\log{\rho} - \log{\sigma}\right)\right],
\end{equation}
such that $S({\rho}\|{\sigma}) \in [0,+\infty)$. Due to its asymmetry and the fact that it is infinite on pure-states, it is not a true metric, however it is nonetheless extremely useful. Defining the REE then follows
\begin{equation}
E_R(\rho) = \min_{\sigma \in \mathcal{D}_{\text{Sep}}} \hspace{-2mm} S(\rho \| \sigma),
\end{equation}
which can be readily employed with respect to parameterised NNS. This can of course generalise to $E_R^{\mathcal{K}}(\sigma)$ given a form of separability. Interestingly, the REE is sub-additive and in general 
\begin{equation}
E_R(\rho\otimes\sigma) \leq E_R(\rho) + E_R(\sigma).
\end{equation}
This lets us define a regularised $n$-shot REE
\begin{equation}
E_R^n(\rho) = \frac{1}{n} \min_{\sigma \in \mathcal{D}_{\text{Sep}}} \hspace{-1mm} S(\rho^{\otimes n} \| \sigma) \leq E_R(\rho).
\end{equation}
The single-shot, standard REE alone is an extremely difficult quantity to compute, largely due to the characterisation of $\mathcal{D}_{\text{Sep}}$ and the unruliness of the QRE. Its computation has recently been explored using an active learning strategy \cite{ActiveLearning}, in which the authors use active learning to compress $\mathcal{D}_{\text{Sep}}$ into a more relevant subset of the separable state space that contributes strongly to the REE. Thanks to the implicit separability of NNS, we may choose an alternative approach where it is possible to optimise some other cost function such as fidelity/trace distance that will simultaneously minimise the QRE towards the optimal REE. In doing so, SNNS should allow for the accurate and efficient approximation of $E_R$, and previously unexplored REEs with respect to other forms of separability $E_R^{\mathcal{K}}$.

\begin{figure}[t!]
\includegraphics[width=\linewidth]{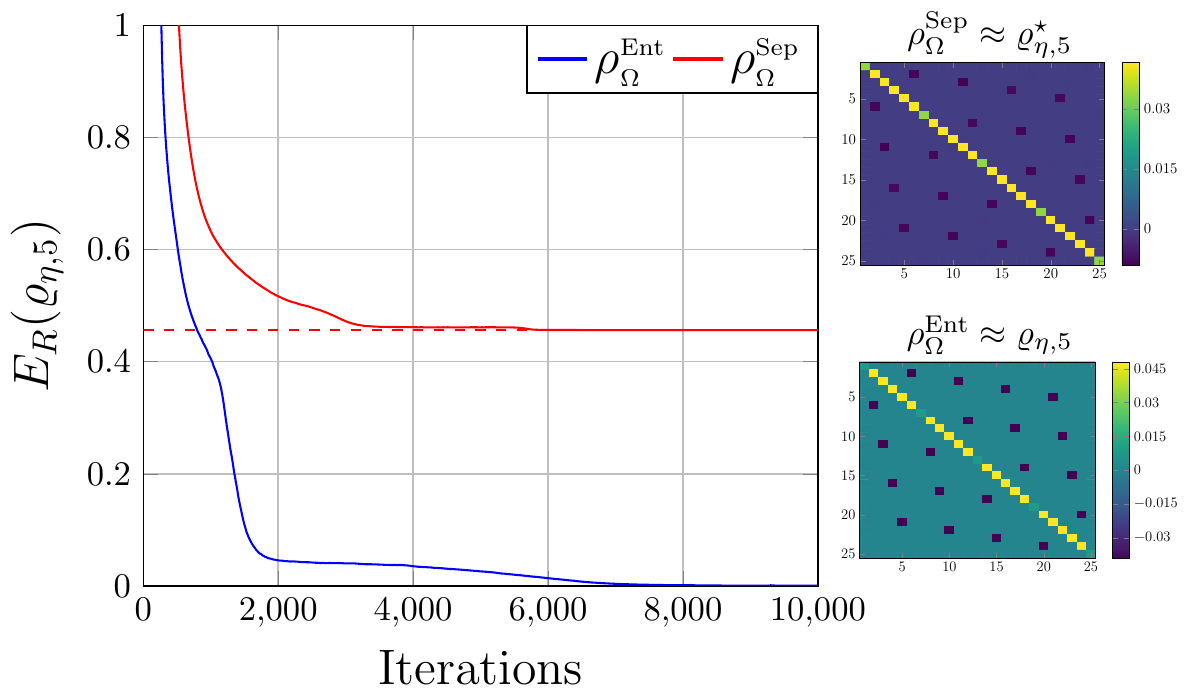}
%0.4564931172906772 | 0.4564355568004036 | 7.534290611732608e-6 (for 5d Werner).
\caption{The classification and entanglement quantification of a $d=5$ Werner state $\varrho_{\eta,d}$, defined in Eq.~(\ref{eq:Werner}) for $\eta = -0.75$. Using NNS, the REE was approximated to within $\epsilon < \!10^{-5}$ precision of the known analytical value $E_R(\varrho_{\eta,d}) \approx 0.4564$ \cite{Vollbrecht}. The entangled network used 10 hidden mixing neurons and 10 hidden pure-state neurons, whilst the separable network used 10 hidden mixing neurons. The density matrices of the (approximate) CSS $\rho_{\Omega}^{\text{Sep}} \approx \varrho_{\eta,5}^{\star}$ and target state approximations are also shown.}
\label{fig:Werner5}
\end{figure}

\section{\label{sec:Results}Applications and Results}

\subsection{Mixed States in $d$-dimensions}
The most substantial generalisation of the methods introduced in Ref.~\cite{ECNNS} is the ability to classify and quantify entanglement in mixed, $d$-dimensional states. To illustrate this improvement, consider the $d$-dimensional Werner state, parameterised by
\begin{equation}
\varrho_{\eta,d} = \frac{(d-\eta)\mathbb{I}_d^{\otimes2} + (d\eta - 1)\mathbb{F}_d}{d(d^2-1)} ,
\label{eq:Werner}
\end{equation}
where $\mathbb{F}_d = \sum_{i,j =0}^{d-1} \ket{ij}\!\bra{ji}$ is the two-qudit flip operator, $\mathbb{I}_d$ is the $d$-dimensional identity operator, and $\eta$ characterises the entanglement properties of the state. For $\eta \in [-1,0]$ the state is entangled, and we can easily quantify this entanglement using the analytically known REE \cite{Vollbrecht},
\begin{equation}
E_R(\varrho_{\eta,d}) = \frac{1+\eta}{2} \log_2(1+\eta) + \frac{1-\eta}{2} \log_2(1-\eta). \label{eq:HWREE}
\end{equation}
In Fig.~\ref{fig:Werner5} we display an optimisation procedure for $d=5, \eta = -0.75$ using an entangled learner $\rho_{\Omega}^{\text{Ent}}$ and a fully separable learner $\rho_{\Omega}^{\text{Sep}}$. The free, entangled learner is able to reconstruct the target Werner state with ease, and an extremely high fidelity, while the fully separable learner correctly classifies the target as entangled. \par
Beyond the obvious entanglement classification, the SNNS is able to quantify the REE of the state, by monitoring the relative entropy $E_R^{\Omega}(\varrho_{\eta,d}) = S(\varrho_{\eta,d}\|\rho_{\Omega}^{\text{Sep}})$ throughout the learning process. As the optimisation converges, $E_R^{\Omega} \rightarrow E_R$, we gather an approximation to the REE of the state. Indeed, under typical optimisation settings, the REE is approximated to within $\epsilon < \!10^{-5}$ precision of the known analytical value $E_R(\varrho_{-0.75,5}) \approx 0.4564$, reinforcing the strength of this approach.

\subsection{Classification of Bound Entangled States}
\begin{figure}[t!]
\includegraphics[width=\linewidth]{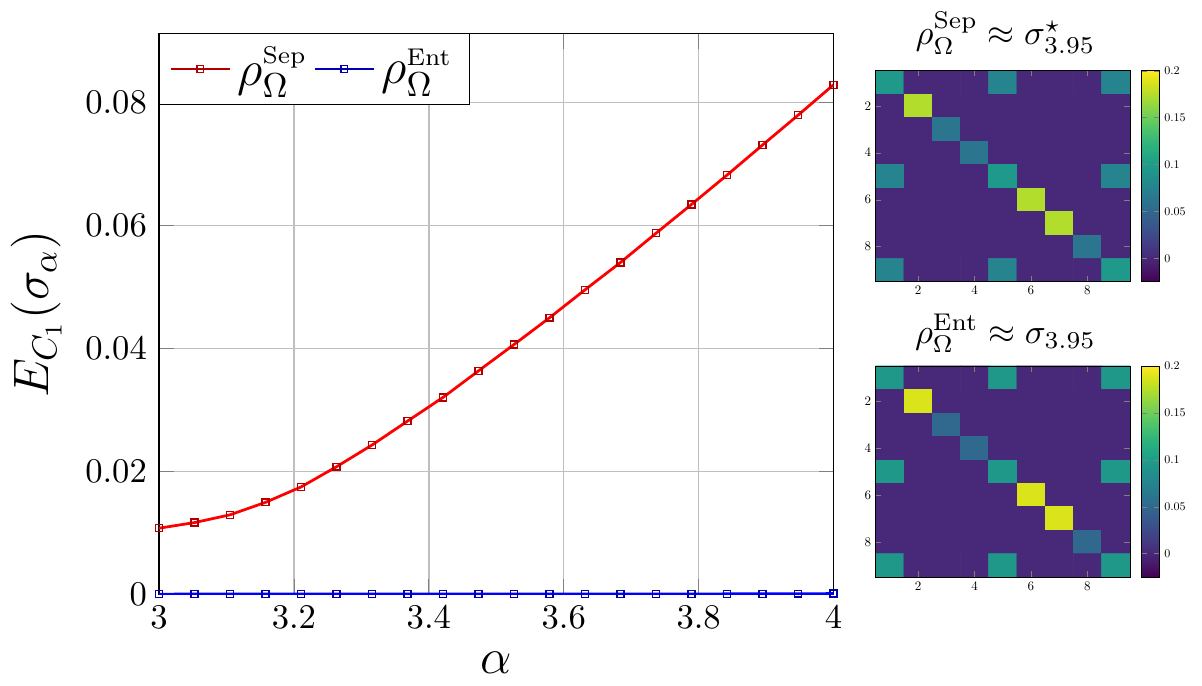}
\caption{Bound entangled state classification. Entangled learners $\rho_{\Omega}^{\text{Ent}}$ (blue) are used to confirm the learnability of the target bound entangled state via NNS. Separable learners $\rho_{\Omega}^{\text{Sep}}$ (red) and then used to classify the target state, and approximate an upper bound on the trace distance from the CSS, $\sigma_{\alpha}^\star$. Here we illustrate density matrices of the approximate CSS, and the target state for $\alpha = 3.95$.}
\label{fig:BoundE}
\end{figure}

The positivity of a partially transposed quantum system \textit{can} be a signature of separability. However it is not universal, and there exist classes of states which are PPT but are entangled, known as bound entangled (BE) states. Here we consider the following two-qutrit state,
\begin{align}
\sigma_+ &= -\frac{1}{3}( \ket{01}\!\bra{01} +  \ket{12}\!\bra{12} +  \ket{20}\!\bra{20}) \nonumber, \\ 
\sigma_- &= \frac{1}{3}( \ket{10}\!\bra{10} +  \ket{21}\!\bra{21} +  \ket{02}\!\bra{02})\nonumber ,\\
\sigma_{\alpha} &= \frac{2}{7} \ket{\Phi^+}\!\bra{\Phi^+} +\frac{\alpha}{7} \sigma_+ + \frac{5-\alpha}{7} \sigma_-,
\label{eq:BE}
\end{align}
where $\ket{\Phi^+} = \frac{1}{\sqrt{3}}(\ket{00} +\ket{11} +\ket{22})$ is a $d=3$ dimensional Bell state. This state is known to satisfy the following entanglement properties \cite{BE}:
\begin{equation}
\sigma_{\alpha} \text{ is } \begin{cases} 
\text{ Separable if  } 2 \leq \alpha \leq 3,\\
\text{ Bound Entangled if  } 3 < \alpha \leq 4,\\
\text{ Free Entangled if  } 4 < \alpha \leq 5.
\end{cases}
\end{equation}
Here we investigate the target state in the bound entangled region, and show that this bipartite state cannot be optimally reconstructed via SNNS. Fig.~\ref{fig:BoundE} depicts the employment of entangled learners $\rho_{\Omega}^{\text{Ent}}$ (blue), and fully separable learners $\rho_{\Omega}^{\text{Sep}}$ (red) to reconstruct $\sigma_{\alpha}$ across the domain $3 < \alpha \leq 4$. \par
For all values of $\alpha$, $\rho_{\Omega}^{\text{Ent}}$ is able to reconstruct the state to a high degree of precision such that the trace distance is $\| \sigma_{\alpha} - \rho_{\Omega}^{\text{Ent}} \|_1 \leq \!10^{-4}$. However, the separable learners are unable to reach this level of reconstruction accuracy. Hence, since $\sigma_{\alpha}$ are learnable via free NNS, the inability of $\rho_{\Omega}^{\text{Sep}}$ to reconstruct $\sigma_{\alpha}$ implies that these states are entangled in this region. Since they are also PPT in this region, we have successfully shown the ability of SNNS to classify bound entanglement.\par
During each constrained optimisation we gather an upper bound on the distance between the target bound entangled state, and its CSS. As said before, this is an upper bound since $\rho_{\Omega}^{\text{Sep}}$ offers an approximation to the CSS, and is potentially loose. Nonetheless the inferred classification is informative. Fig.~\ref{fig:BoundE} plots the trace distance $\| \sigma_{\alpha} - \rho_{\Omega}^{\text{Sep}} \|_1$, shown to steadily rise as $\alpha$ increases, which is expected as $\sigma_{\alpha}$ becomes freely entangled for $4 < \alpha \leq 5$.

\subsection{Detection and Measurement of Multipartite Entanglement}
The versatility of the $\mathcal{K}$-separable state design means that we can explore entanglement classification and quantification methods that are otherwise very difficult. In particular, we may construct a NNS protocol that is able to witness W/GHZ-state entanglement, and measure W/GHZ-type correlations in both pure and mixed quantum states. Consider the three-qubit W and GHZ states respectively \cite{Wstate,GHZstate}
\begin{align*}
&\ket{\text{W}} = \frac{1}{\sqrt{3}} \left( \ket{001} + \ket{010} + \ket{100} \right), \\
&\ket{\text{GHZ}} = \frac{1}{\sqrt{2}} \left( \ket{000} + \ket{111}\right).
\end{align*}
These are both maximally entangled three party states. However they possess two inequivalent forms of tripartite entanglement, such that  $\ket{\text{W}}$ cannot be transformed into $\ket{\text{GHZ}}$ by means of LOCC (local operations and classical communications) strategies. The key difference in these forms of entanglement is their \textit{robustness} i.e.~when a party is removed from a\text{ GHZ }state the remaining states are separable, whilst a \text{W-}state remains entangled. Therefore a \text{W-}state possesses strict bipartite entanglement between all three parties, whereas\text{ GHZ }entanglement can be achieved via ``relayed entanglement" \footnote{We refer to relayed entanglement as that which is caused indirectly through mutually entangled parties. For example, if $\mathcal{K} = \{1,2 | 2,3\}$, entanglement is indirect between parties 1 and 3 - it is \textit{relayed}.}.\par
\begin{figure*}[t!]
 (a) Classification of $\ket{W}$, \hspace{2.5cm} (b) $\sigma_{\text{\tiny W}}^{p}$/$\sigma_{\text{\tiny GHZ}}^{p}$ for $p=\frac{1}{3}$, \hspace{2.5cm} (c) REE for $\mathcal{E}_{\text{\tiny D}}(\sigma_W^p)$.\\ 
\hspace{-5mm}\includegraphics[width=0.334\linewidth]{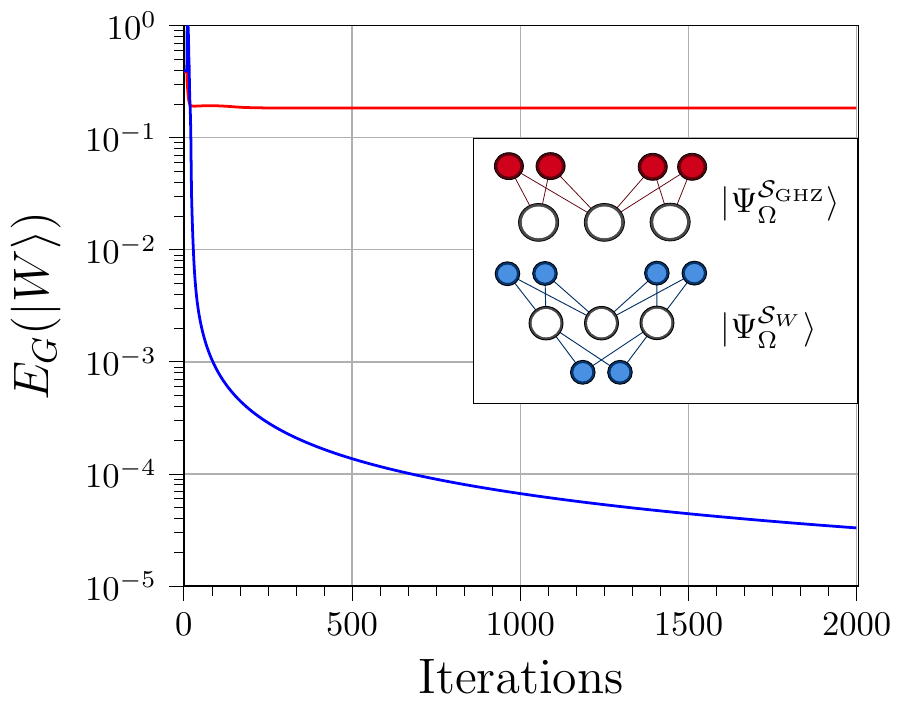}
\includegraphics[width=0.33\linewidth]{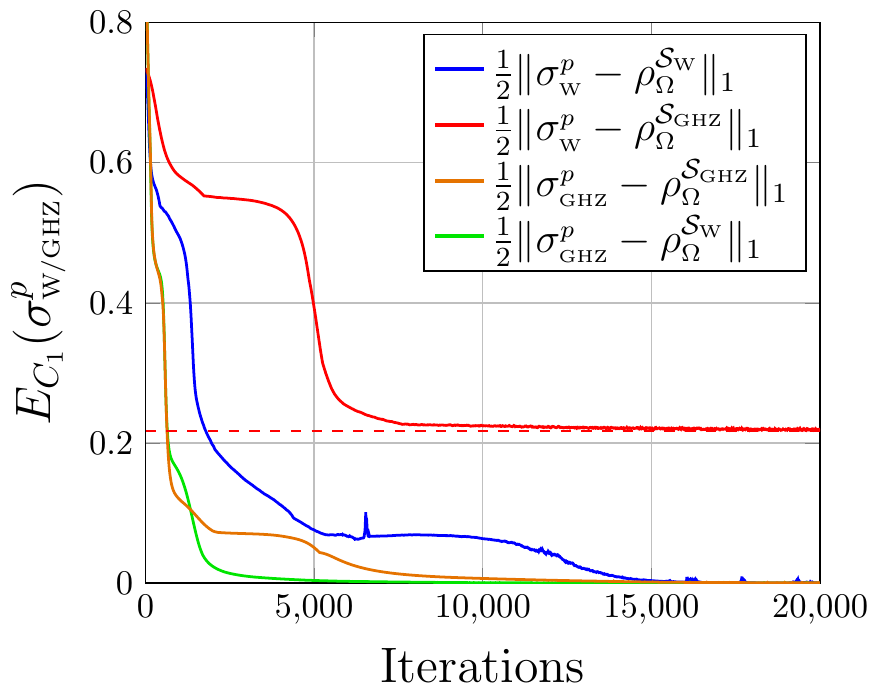}
\includegraphics[width=0.32\linewidth]{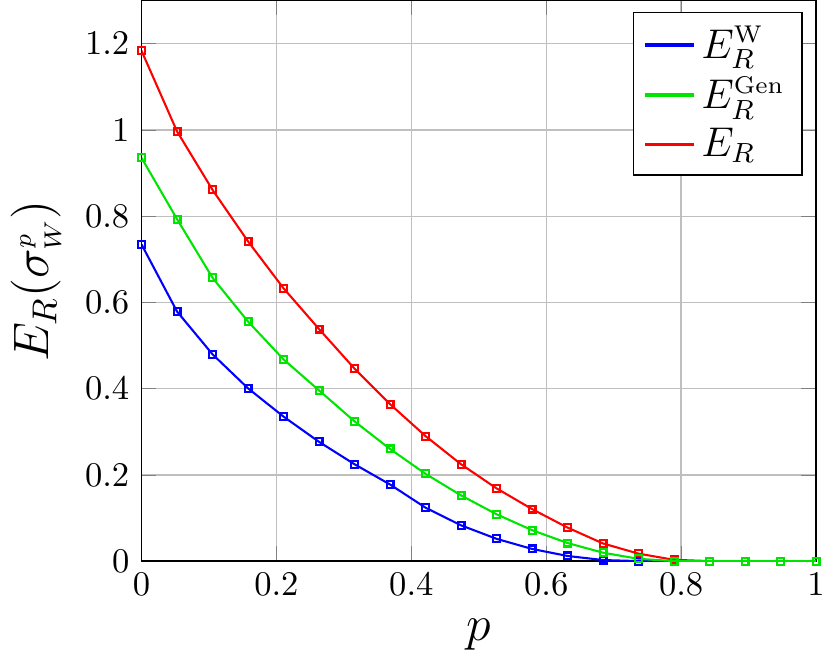}
\caption{Classification and quantification of $d=2$ W/GHZ type entanglement using NNS. Panel (a) shows the classification of W-type entanglement using two NNS designed according to the partition sets $\mathcal{K}_{\text{GHZ}} = \{1,2|2,3\}$ and $\mathcal{K}_{\text{W}} = \{1,2|2,3|1,3\}$. If a variational state endowed with $\mathcal{K}_{\text{W}}$-separability can optimally reconstruct a target that $\mathcal{K}_{\text{GHZ}}$ cannot, then it must possess W-type entanglement. In turn, we locate the closest GHZ-entangled state to $\ket{W}$. In Panel (b) this is extended to mixed, depolarised W/GHZ-states for $p = \frac{1}{3}$. Panel (c) depicts different versions of the REE upper bounds on a depolarised W-state $\sigma_{\text{W}}^p$ with respect to depolarising probability. Here we plot three types of REE: The fully separable REE $E_R$ (red), the genuine tripartite REE $E_R^{\text{Gen}}$ (green) and the strictly W-type entanglement REE $E_R^{\text{W}}$ (blue).
}
\label{fig:WtypeEnt1}
\end{figure*}
To classify between these states, we must define a partition set that is capable of capturing\text{ GHZ }correlations, but incompletely capture \text{W-}type correlations. The non-disjoint separability set 
\begin{equation}
\mathcal{K}_{\text{W}} = \{1,2|2,3|1,3\}, 
\end{equation}
is capable of learning both\text{ W }and\text{ GHZ }entangled states, as it strictly specifies bipartite entanglement between all parties. However, one can construct the partition set
\begin{equation}
\mathcal{K}_{\text{GHZ}} = \{i, j| i ,k\}, i \neq j \neq k \in \{1,2,3\}, \label{eq:GHZSep}
\end{equation}
which is any possible permutation of two subsets of $\mathcal{K}_{\text{W}}$. Programming a NNS according to $\mathcal{K}_{\text{GHZ}}$ does not allow the network to capture direct correlations between qubits $j$ and $k$, and will therefore provide an insufficient ansatz to reconstruct W-states. This forms a witness for W-type entanglement; if a target state is learnable via a NNS endowed with $\mathcal{K}_{\text{W}}$-separability, but is not learnable via $\mathcal{K}_{\text{GHZ}}$-separability, then the state is verified as possessing W-type entanglement. Furthermore, by constructing entanglement measures $E_{\Omega}^{\mathcal{K}_{\text{GHZ}}}$ we are able to measure the amount of W-type correlations within a target state. \par
Figure~\ref{fig:WtypeEnt1}(a) shows the pure-state classification of a three-qubit W-state, where the non-disjoint network architectures perform classification easily. Note that these three-qubit partitions can be analogously embedded into larger, $n$-qudit systems in order to study more complex forms of entanglement.\par

Realistically, multipartite entangled resources for future quantum communication/computing protocols will be noisy and imperfect. Generating and distributing multipartite entanglement over noisy quantum channels is fundamental for many future quantum technologies, particularly for secure communications  and quantum networks \cite{AdvCrypt,End2End,MultiCastBs,GenConf,SP_Conf,ConfRev,WehnerWStates,Anon_QN}. Therefore it is a more interesting challenge to consider the classification and quantification of tripartite entanglement subject to decoherence. For instance, one can consider versions of $\ket{\text{W}}$/$\ket{\text{GHZ}}$ in which each qudit has been passed through a depolarising channel
\begin{equation}
\mathcal{E}_{\text{\tiny D}}(\rho) = (1-p)\rho + \frac{p}{d^n}\mathbb{I}_d^{\otimes n}, \label{eq:DepolCh}
\end{equation}
where $n$ denotes the number of qudits being acted on (in this case $n=3$). We denote these noisy, three-qubit states as
\begin{align}
\sigma_{\text{W}}^p &= (1-p)\ket{\text{W}}\!\bra{\text{W}} + \frac{p}{8}\mathbb{I}_2^{\otimes 3},\\
\sigma_{\text{GHZ}}^p &= (1-p)\ket{\text{GHZ}}\!\bra{\text{GHZ}} + \frac{p}{8}\mathbb{I}_2^{\otimes 3}.
\end{align}
Using mixed NNS programmed with different separabilities, we may then easily distinguish between the entanglement properties of noisy W/GHZ-states subject to depolarising channels. Indeed, Fig.~\ref{fig:WtypeEnt1}(b) shows that for $p = \frac{1}{3}$ we can perform this classification. Given two learners $\rho_{\Omega}^{\mathcal{K}_\text{\tiny W}}$ and $\rho_{\Omega}^{\mathcal{K}_\text{\tiny GHZ}}$, it is clear that both are able to optimally reconstruct the noisy GHZ-state, whilst only $\rho_{\Omega}^{\mathcal{K}_{\text{W}}}$ is able to optimally reconstruct the noisy W-state, completing the classification.\par
This is taken a step further in Fig.~\ref{fig:WtypeEnt1}(c) where different versions of the REE of $\sigma_{\text{W}}^p$ is monitored for various depolarising probabilities. This plot describes three forms of REE: 
\begin{itemize}
\item The standard $E_R$ (red) defined on the space of all fully separable states (using the partition set $\mathcal{K}_{\text{FS}} = \{1|2|3\}$) which measures the amount of any entanglement present.
\item The genuine tripartite entangled REE, $E_R^{\text{Gen}}$ (green), using the bi-separable partition sets $\mathcal{K}_{\text{BS}} = \{i,j|k\}, i\neq j\neq k \in \{1,2,3\}$, which measures the amount of genuine tripartite entanglement in the state (W or GHZ correlations).
\item The W-REE, $E_R^{\text{W}}$ (blue) using the partition set $\mathcal{K}_{\text{GHZ}}$ in Eq.~(\ref{eq:GHZSep}), which measures the amount of genuine, tripartite, strictly W-type entanglement within the state.
\end{itemize}
By employing more complex separable architectures, we may study how different forms of entanglement behave with respect to environmental properties, such as depolarisation. By measuring $E_R^{\text{Gen}}$ and $E_R^{\text{W}}$ for instance, we may monitor the decoherence of genuine tripartite entanglement, rather than \textit{any} entanglement as done so by $E_R$. Such characterisations could prove very useful in communication/networking scenarios, where genuine multipartite entanglement is critical to performance.\par
It is important to remind the reader that these are upper bounds. The standard REE upper bound is expected to be tight, as fully separable NNS architectures precisely capture full separability. However, $\mathcal{K}_{\text{BS}}$ and $\mathcal{K}_{\text{GHZ}}$ are degenerate, e.g. $\mathcal{K}_{\text{BS}} =  \{i,j|k\}$ has 3 unique forms. Since mixed SNNS are restricted to consistent separabilities, there may be convex combinations of states of these separabilities that produce tighter bounds. It is unknown if this is the case, nonetheless $E_R^{\text{Gen}}$ and $E_R^{\text{W}}$ provide informative upper bounds on these unique entanglement measures.

\begin{figure}[t!]
\hspace{4mm} (a) Depolarising Channel \\
\includegraphics[width=0.9\linewidth]{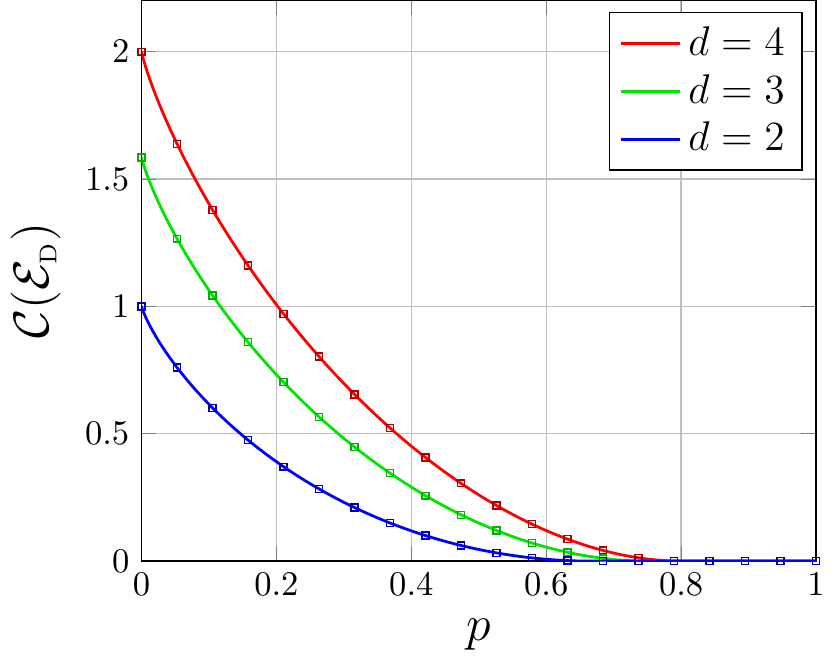} \\
\hspace{1cm} \\
\hspace{4mm} (b) Holevo Werner Channel \\
\includegraphics[width=0.9\linewidth]{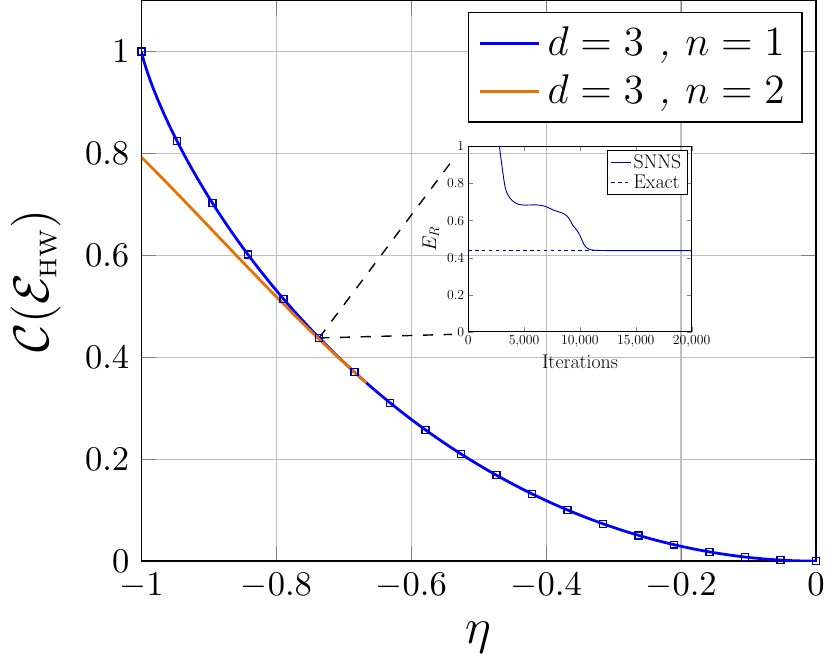}
\caption{PLOB channel capacity upper bounds computed via separable neural network states. Continuous plots are exact, while the scatter plots are SNNS data. Panel (a) displays the communication capacities for $d=2,3,4$ dimensional quantum systems in a depolarising channel of depolarising probability $p$, using mixed, qudit SNNS ansatzes. Panel (b) depicts the capacity for Holevo-Werner (HW) qutrit channels. The network states approximate the REE to a typical accuracy of $\epsilon < \!10^{-5}$, hence reproducing the capacities to a very high degree of precision.}
\label{fig:Bounds}
\end{figure}

\subsection{Ultimate Limits for Channel Capacities}
We may provide a more practical example for the use of SNNS in the realm of quantum communications, using them to approximate upper bounds of quantum channel capacities. Introduced in Ref.~\cite{PLOB}, the Pirandola-Laurenza-Ottaviani-Banchi (PLOB) bound is an ultimate upper bound on the two-way assisted quantum (and secret-key) capacity $\mathcal{C}(\mathcal{E})$ for a given quantum channel $\mathcal{E}$. Its derivation is based on the techniques of channel simulation and teleportation stretching, which have proven to be extremely versatile in a number of settings \cite{TCS,AdaptiveNoiseEst,End2End,DenseCodingCaps,Laurenza2018,FLQCD,ConvBanchi}. An essential class of quantum channels are those which are teleportation covariant, meaning that they satisfy the condition
\begin{equation}
\mathcal{E}(U \rho U^{\dagger}) = V\mathcal{E}(\rho)V^{\dagger},
\end{equation}
for some pair of teleportation unitaries $\{U,V\}$. Let us define the Choi matrix of a $d$-dimensional channel $\mathcal{E}$ as the result of passing one mode of a maximally entangled state $\Phi^{+}$ through the $\mathcal{E}$, and the other through an identity channel $\mathcal{I}$
\begin{equation}
\rho_{\mathcal{E}} =  \mathcal{I}\otimes\mathcal{E} [{\Phi^{+}}],
\end{equation}
where the maximally entangled state may take the form $\Phi^{+} = \frac{1}{d} \sum_{i,j = 0}^{d-1} \ket{ii}\!\bra{jj}$. For teleportation covariant channels, the ultimate channel capacity can then be upper bounded in a remarkably simple way \cite{PLOB}
\begin{equation}
\mathcal{C}(\mathcal{E}) \leq E_{R}^n(\rho_{\mathcal{E}}) \leq  E_{R}(\rho_{\mathcal{E}}) ,
\end{equation}
where $E_R$ is the standard relative entropy of entanglement (and $E_R^n$ its $n$-shot version). SNNS can be used to approximate upper bounds on these channel capacities, via constrained reconstruction of the Choi state of the desired quantum channel.\par 
We consider two important, teleportation covariant, $d$-dimensional quantum channels in an effort to illustrate the effectiveness of our approach: The depolarising channel considered in Eq.~(\ref{eq:DepolCh}), and the Holevo-Werner channel \cite{HWChannels,HWCope,HWCope2}. The Choi states of these channels are the classes of isotropic states and Werner states respectively, whose REE bounds are known analytically. Therefore, we can compare the numerical performance of computing the REE via SNNS with the known, exact bounds.\par
Fig.~\ref{fig:Bounds}(a) reports REE bounds on the capacity of depolarising channels for dimensions $d=2,3,4$. Approximating these bounds via separable network states requires the targeted reconstruction of the isotropic state,
\begin{equation}
\rho_{\mathcal{E}_{\text{\tiny D}}} = (1-p)\Phi^{+}+ \frac{p}{d^2} \mathbb{I}_d^{\otimes2}.
\end{equation}
Using a bipartite SNNS $\rho_{\Omega}^{\text{Sep}}$, and attempting to learn the target Choi state leads to an approximation of the REE of said state. Performing this optimisation for many depolarising probabilities $p$, the results in Fig.~\ref{fig:Bounds}(a) can be produced. This is be achieved to a very high degree of accuracy, reproducing the analytical bounds with an average error $\sim \epsilon < \!10^{-5}$. Furthermore, these bounds can be computed very efficiently by performing each optimisation sequentially, initialising the network parameters using the results of previous optimisations \footnote{A scenario in which efficiency can be greatly enhanced, is the the study of evolving states. Consider the results from Fig.~\ref{fig:BoundE}-\ref{fig:Bounds}. In a number of instances, we are classifying/quantifying the entanglement of a target state which is changing incrementally (and by a small amount) throughout an interval. Consider a NNS $\rho_{\Omega}$ that learns a state $\sigma$. It is logical to assume that if the target state is evolved by some small amount, $\sigma^{\prime} = \sigma + \delta \sigma$, the network $\Omega$ will only need to be optimised by a small amount $\Omega^{\prime} = \Omega + \delta \Omega$. Therefore, when studying evolving target states, it is extremely useful to initialise each state using the parameter distribution of the previous learner. This not only simplifies learning and performance, but increases efficiency dramatically; the initial target can be reconstructed over a number of optimisation steps $S$, but subsequent alterations to the network only require a fraction of $S$ steps.
}.\par
In Fig.~\ref{fig:Bounds}(b) we give REE upper bounds for the HW channel, which takes the form
\begin{equation}
\mathcal{E}_{\text{\tiny HW}}^{\eta,d} (\rho) = \frac{ (d-\eta)\mathbb{I}_d^{\otimes 2} + (d\eta - 1)\rho^{{T}}}{d^2 - 1},
\end{equation}
such that $T$ superscript denotes the transposition. The Choi state of the HW channel is the $d$-dimensional Werner state, introduced in Eq.~(\ref{eq:Werner}). The single shot REE bounds for the HW channel are analytically known and given in Eq.~(\ref{eq:HWREE}), and are independent of dimension $d$. Again, this single shot bound is approximated to a good precision, as shown in the results.\par
For Werner states of dimension $d > 2$, their REE is known to be strictly sub-additive when $\eta < -\frac{d}{2}$, and previous studies have explored the two-shot REE for these Choi states \cite{HWCope}, which can therefore be used to tighten these upper bounds. For instance, in Fig.~\ref{fig:Bounds}(b) the two-shot capacity can be seen to significantly tighten the bounds for $d=3$. In order to compute these tighter bounds, one must modify the definition of the $n$-shot quantities slightly. Now the minimisation is performed with respect to the space of all locally bi-separable states. Consider the $n$-copy Werner state, and let us label each copy with indices of its modes $\{i,j\}$,
\begin{equation}
\varrho_{\eta,d}^{\otimes n} = \varrho_{\eta,d}^{\{1,2\}} \otimes \varrho_{\eta,d}^{\{3,4\}} \otimes \ldots \otimes \varrho_{\eta,d}^{\{2n-1,2n\}}.
\end{equation}
The goal is now to find the CSS that possesses the following bi-separability
\begin{align}
\sigma^{n} =  \sigma_a^{\{1,3,5,\ldots,2n-1\}} \otimes \sigma_b^{\{2,4,6,\ldots, 2n\}},
\end{align}
where we have permuted the labels into a bi-separable decomposition such that each state belongs to exclusively even or odd mode labels. This corresponds to a situation where two users each possess $n$ local modes, and their goal is to produce the closest state to $\varrho_{\eta,d}^{\otimes n}$ that is bi-separable between them. In general this is a very difficult task, and while beyond the scope of this paper, poses as an interesting future application for SNNS.

\section{Conclusions and Outlook\label{sec:Conclusions}}
We have generalised the concept of NNS with programmable separability to mixed, $d$-dimensional quantum states. We discussed a number of neural network architectures for the description of quantum states, and detailed how their entanglement properties may be controlled via constraints placed on network connectivity. It was shown that network connectivity controls entanglement structure on a very specific level, requiring distinctions between certain forms of entanglement. Outlining one of many possible optimisation protocols, methods of classification and quantification via SNNS have been logically developed, and applied in a number of important settings. We then studied a practical application of these tools in the bounding of ultimate quantum channel capacities, showing that they can reproduce the PLOB bounds for DV channels with high precision.\par
There are a number of valuable future directions in which SNNS may be explored and expanded. While an optimisation scheme based on the vectorised fidelity is effective for a variety of applications (as shown in this work) more sophisticated optimisation protocols could enhance performance for more specific entanglement measures. In particular, a gradient descent method that directly minimises the relative entropy (or some variant thereof) would provide a more effective computation of the REE for complex states. This would also lend well to the study of $n$-shot REE quantities with applications in quantum channel capacities, and the characterisation of more complex bound entangled states (such as those constructed from un-extendible product bases). Combining these tools with those from practical quantum tomography could also be extremely useful, e.g.~where SNNS may be used to certify the effectiveness an entanglement distribution protocol.
\begin{acknowledgements}
C.H acknowledges funding from the EPSRC via a Doctoral Training Partnership (EP/R513386/1). MP acknowledges the H2020-FETOPEN-2018-2020 project TEQ (grant nr.~766900), the DfE-SFI Investigator Programme (grant 15/IA/2864), COST Action CA15220, the Royal Society Wolfson Research Fellowship (RSWF\textbackslash~R3\textbackslash183013), the Leverhulme Trust Research Project Grant (grant nr.~RGP-2018-266), the UK EPSRC (grant nr.~EP/T028106/1). SP acknowledges funding from the European Union’s Horizon 2020 Research and Innovation Action under grant agreement No.~862644 (Quantum readout techniques and technologies, QUARTET).
\end{acknowledgements}

%\bibliography{ECNNS}


\appendix

\section{\label{sec:DerivsULM}Learning with Complex-Exponential Ansatz for Mixed States}
As discussed in Section \ref{sec:MixedStates}, one can make use of a restructuring of the mixed state ansatz into complex exponential form in order to take better control of the learning procedure. Indeed, the total mixed state can be expressed as
\begin{align}
\rho_{\text{ $\Omega,\Pi,\Xi$}}^{\bs{\alpha},\bs{\beta}} = e^ {i\log \left({\Phi_{\Xi}(\bs{\alpha},\bs{\beta}) \vartheta_{\Omega}(\bs{\alpha},\bs{\beta})}\right)} \Gamma_{\Pi}(\bs{\alpha},\bs{\beta}) r_\Omega(\bs{\alpha},\bs{\beta}),
\end{align}
such that the state is constructed from three variational parameter sets, where $r_\Omega$ and $\Gamma_{\Pi}$ assume responsibility for the magnitude of any element of the density-matrix, while functions $\Phi_{\Xi}$ and $\vartheta_{\Omega}$ are responsible for the complex phase of such elements. Consider a target state $\chi$ which also admits the following decomposition
\begin{equation}
\chi^{\bs{\alpha},\bs{\beta}} = \lambda(\bs{\alpha},\bs{\beta}) e^{i \log{ \xi(\bs{\alpha},\bs{\beta})}}.
\end{equation}\par
The pure density-matrix phase/amplitude functions $\Phi_{\Xi}$ and $\Gamma_{\Pi}$ respectively, are parameterised by real valued parameter sets. Furthermore, they are decomposed with respect to their pure-state wavefunctions, as shown in Eq.~(\ref{eq:PureDecomp}). The logarithmic derivatives of the pair of pure-state phase functions take the form
\begin{align}
& \frac{\partial \log \ket{\Phi_{\Xi}}}{\partial \Xi_k} = \sum_{\bs{\alpha},\bs{\beta}} \left( \frac{\partial \log \varphi(\bs{\alpha})}{\partial \Xi_k}  - \frac{\partial \log \varphi(\bs{\beta})}{\partial \Xi_k} \right) \label{eq:derivX},
\end{align}
while the amplitude function derivatives are
\begin{align}
&\frac{\partial \log \ket{\Gamma_{\Pi}}}{\partial \Pi_k} = \sum_{\bs{\alpha},\bs{\beta}} \left( \frac{\partial \log \sigma(\bs{\alpha})}{\partial \Pi_k}  + \frac{\partial \log \sigma(\bs{\beta})}{\partial \Pi_k} \right) \label{eq:derivP}.
\end{align}
Meanwhile, the mixing state phase/amplitude wavefunctions $\vartheta_{\Omega}$ and $r_{\Omega}$ respectively are based on complex parameters. In this case, it is expedient to take derivatives with respect to real and imaginary components, i.e.~$\frac{\partial \log \ket{r_{\Omega}}}{\partial \text{Re}(\Omega_k)}$, $\frac{\partial \log \ket{r_{\Omega}}}{\partial \text{Im}(\Omega_k)}$, $\frac{\partial \log \ket{\vartheta_{\Omega}}}{\partial \text{Re}(\Omega_k)}$ and $\frac{\partial \log \ket{\vartheta_{\Omega}}}{\partial \text{Re}(\Omega_k)}$ which can be treated separately. All these derivatives take real, compact and easily derived forms with respect to the neural network parameters, making gradient computations straightforward.\par
The learning procedure of minimising the negative logarithmic fidelity between a target vectorised density-matrix $\ket{\chi}$ and the mixed NNS is given by the usual update rule in Section \ref{sec:Learn_Cl_Q}. Defining the quantity
\begin{equation}
\Delta( {\bs{\alpha},\bs{\beta}} ) ={{\braket{\rho_{\Omega,\Pi,\Xi} | \chi}}}^{-1} {e^{i \log\frac{{\Phi_{\Xi}}(\bs{\alpha},\bs{\beta})\vartheta_{\Omega} (\bs{\alpha},\bs{\beta}) }{\xi(\bs{\alpha},\bs{\beta})}}},
\end{equation}
where ${\braket{\rho_{\Omega,\Pi,\Xi} | \chi}}$ is the vectorised overlap between the variational and target state, we can then make use of the following gradients,
\begin{widetext}
\begin{align}
&\nabla_k^{\Gamma_\Pi}\mathcal{L} = \sum_{\bs{\alpha},\bs{\beta}} \Bigg[\frac{ r_{\Omega}^2(\bs{\alpha},\bs{\beta})  {\Gamma_\Pi}(\bs{\alpha},\bs{\beta})}{|\rho_{\Omega,\Pi,\Xi}|^2} - \lambda(\bs{\alpha},\bs{\beta}) \> r_{\Omega}(\bs{\alpha},\bs{\beta}) \> \text{Re}\left[ \Delta( {\bs{\alpha},\bs{\beta}} )
\right] \Bigg] \cdot {\mathcal{O}}_{k}^{\Pi} \ket{\Gamma_\Pi}, \\
&\nabla_k^{r_\Omega}\mathcal{L} =\sum_{\bs{\alpha},\bs{\beta}} \Bigg[ \frac{ {\Gamma_\Pi^2}(\bs{\alpha},\bs{\beta}) r_{\Omega}(\bs{\alpha},\bs{\beta})}{|\rho_{\Omega,\Pi,\Xi}|^2} - \lambda(\bs{\alpha},\bs{\beta}) \> \Gamma_{\Pi}(\bs{\alpha},\bs{\beta}) \> \text{Re}\left[ \Delta( {\bs{\alpha},\bs{\beta}} )
\right] \Bigg] \cdot {\mathcal{O}}_{k}^{\Omega_r} \ket{r_\Omega}, \\
&\nabla_k^{\Phi_\Xi}\mathcal{L} = - \sum_{\bs{\alpha},\bs{\beta}} \Bigg[ \frac{ r_{\Omega} (\bs{\alpha},\bs{\beta}) \lambda(\bs{\alpha},\bs{\beta}) \> {\Gamma_\Pi}(\bs{\alpha},\bs{\beta})}{{\Phi_\Xi}(\bs{\alpha},\bs{\beta})} \> \text{Im}\left[
\Delta( {\bs{\alpha},\bs{\beta}} )
\right]\Bigg] \cdot {\mathcal{O}}_{k}^{\Xi} \ket{\Phi_\Xi},\\
&\nabla_k^{\vartheta_\Omega}\mathcal{L} = - \sum_{\bs{\alpha},\bs{\beta}} \Bigg[ \frac{ r_{\Omega} (\bs{\alpha},\bs{\beta}) \lambda(\bs{\alpha},\bs{\beta}) \> {\Gamma_\Pi}(\bs{\alpha},\bs{\beta})}{{\vartheta_\Omega}(\bs{\alpha},\bs{\beta})} \> \text{Im}\left[
\Delta( {\bs{\alpha},\bs{\beta}} )
\right]\Bigg] \cdot {\mathcal{O}}_{k}^{\Omega_{\vartheta}} \ket{\vartheta_\Omega}.
\end{align}
\end{widetext}
Here, $|\rho_{\Omega,\Pi,\Xi}|^2$ is the magnitude of the vectorised density-matrix. Furthermore ${\mathcal{O}}_{k}^{{\Omega_r}} = \text{diag} \left({\partial_{{\Omega_k}} \log \ket{r_\Omega}}\right) $ and ${\mathcal{O}}_{k}^{{\Omega_\vartheta}} = \text{diag} \left({\partial_{{\Omega_k}} \log \ket{\vartheta_\Omega}}\right) $ are the diagonal matrices with mixing layer gradients. Again, these are treated separately with respect to real and imaginary valued parameters in $\Omega$.

\end{document}